\documentclass[apj]{emulateapj}
\usepackage{apjfonts}

\begin{document}

\slugcomment{Accepted to the Astrophysical Journal [2009, ApJ, 694, 1281] Received: 2008 August 14}

\title{GALEX--SDSS Catalogs for Statistical Studies}

\author{Tam\'as Budav\'ari\altaffilmark{1},
S\'ebastien Heinis\altaffilmark{1},
Alexander S. Szalay\altaffilmark{1},
Mar\'{\i}a Nieto-Santisteban\altaffilmark{1},
Jayant Gupchup\altaffilmark{2},
Bernie Shiao\altaffilmark{3},
Myron Smith\altaffilmark{3},
Ruixiang Chang\altaffilmark{4},
Guinevere Kauffmann\altaffilmark{5},
Patrick Morrissey\altaffilmark{6},
David Schiminovich\altaffilmark{7},
Bruno Milliard\altaffilmark{8},
Ted K. Wyder\altaffilmark{6},
D. Christopher Martin\altaffilmark{6},
Tom A. Barlow\altaffilmark{6},
Mark Seibert\altaffilmark{9},
Karl Forster\altaffilmark{6},
Luciana Bianchi\altaffilmark{1},
Jose Donas\altaffilmark{8},
Peter G. Friedman\altaffilmark{6},
Timothy M. Heckman\altaffilmark{1},
Young-Wook Lee\altaffilmark{10},
Barry F. Madore\altaffilmark{9},
Susan G. Neff\altaffilmark{11},
R. Michael Rich\altaffilmark{12},
and
Barry Y. Welsh\altaffilmark{13}
}

\shortauthors{BUDAV\'ARI ET~AL.}
\email{budavari@jhu.edu}

\altaffiltext{1}{Department of Physics \& Astronomy, The Johns Hopkins
University, 3701 San Martin Dr., Baltimore, MD 21218}

\altaffiltext{2}{Department of Computer Science, The Johns Hopkins
University, 3400 N. Charles St., Baltimore, MD 21218}

\altaffiltext{3}{Space Telescope Science Institute, 3700 San Martin Dr., Baltimore, MD 21218}

\altaffiltext{4}{Shanghai Astronomical Observatory, 80 Nandan Road, Shanghai 200030, China}

\altaffiltext{5}{Max-Planck-Institute f\"ur Astrophysik,
Karl-Schwarzschild-Strasse 1, 85748 Garching, Germany}

\altaffiltext{6}{California Institute of Technology, MC 405-47, 1200
E. California Blvd., Pasadena, CA 91125}

\altaffiltext{7}{Department of Astronomy, Columbia University, New
York, NY 10027, USA}

\altaffiltext{8}{Laboratoire d'Astrophysique de Marseille, BP 8, Traverse
du Siphon, 13376 Marseille Cedex 12, France}

\altaffiltext{9}{Observatories of the Carnegie Institution of Washington,
813 Santa Barbara St., Pasadena, CA 91101}

\altaffiltext{10}{Center for Space Astrophysics, Yonsei University, Seoul
120-749, Korea}

\altaffiltext{11}{Laboratory for Astronomy and Solar Physics, NASA Goddard
Space Flight Center, Greenbelt, MD 20771}

\altaffiltext{12}{Department of Physics and Astronomy, University of
California, Los Angeles, CA 90095}

\altaffiltext{13}{Space Sciences Laboratory, University of California at
Berkeley, 601 Campbell Hall, Berkeley, CA 94720}

\shorttitle{GALEX--SDSS Catalogs for Statistical Studies}

\begin{abstract}
We present a detailed study of the Galaxy Evolution Explorer's
photometric catalogs with special focus on the statistical properties
of the All-sky and Medium Imaging Surveys. We introduce the concept of
primaries to resolve the issue of multiple detections and follow a
geometric approach to define clean catalogs with well-understood
selection functions. We cross-identify the GALEX sources (GR2+3) with Sloan
Digital Sky Survey (DR6) observations, which indirectly provides an
invaluable insight about the astrometric model of the UV sources and
allows us to revise the band merging strategy. We derive the formal
description of the GALEX footprints as well as their intersections
with the SDSS coverage along with analytic calculations of their areal
coverage. The crossmatch catalogs are made available for the
public. We conclude by illustrating the implementation of typical
selection criteria in SQL for catalog subsets geared toward statistical
analyses, e.g., correlation and luminosity function studies.
\end{abstract}

\keywords{catalogs -- methods: statistical -- surveys -- ultraviolet: general}

\section{Introduction}
\label{sec:intro}

The ultraviolet (UV) range of the spectrum is a tracer of recent star
formation within galaxies \citep[e.g.,][]{Kennicutt_1998}. It has been
intensively used \citep[e.g.,][]{Giavalisco_2002} at high redshifts to
study the properties of galaxies selected from the Lyman Break
technique \citep{Steidel_1995}. As the restframe UV light at $z
\lesssim 1$ is not observable from Earth, the UV properties of objects
have been better known in the distant Universe than in
the local Universe. Since the launch of the Galaxy Evolution Explorer
\citep[GALEX;][]{galex}, a NASA small explorer satellite designed and
built to image the sky in the ultraviolet at $z<1$, a new window has
been opened to connect low and high redshift UV
observations. UV data are of primary interest for studying star
formation over timescales of about 100 Myrs, and low redshift UV data
are useful to interpret similar high redshift data. The GALEX
measurements themselves, however, do not provide enough information
for most studies due to the lack of angular resolution
and narrow spectral coverage, which do not allow for a reliable
star--galaxy separation. The single color of the two bands is also
a serious restriction on the potential applications. The solution is
to cross-identify the GALEX sources to other catalogs, in particular to the Sloan
Digital Sky Survey \citep[SDSS;][]{york}. Most GALEX observations are designed to cover
regions of the sky already observed by the SDSS at a comparable depth.

In \citet{seibert05} and \citet{bianchi05,bianchi07} we have discussed various
aspects of the associated catalogs for preliminary data releases. 
Existing GALEX catalogs have a number of shortcomings that include the non-uniqueness
of sources detected and multiple fields. Building on the experience from
a series of previous studies,
we systematically analyze the issues of GALEX catalog creation in conjunction with
the SDSS datasets to define a clean sample optimized for statistical studies.
The impact of this new compilation is most significant on applications that
require a good understanding of the selection effects and rely on the
knowledge of the precise coverage of the survey, e.g., clustering and
luminosity function studies.

The structure of the paper is as follows.
First we look at the GALEX catalogs in \S~\ref{sec:galex},
and provide solutions for common issues such as multiple observations
of the same sources.
In \S~\ref{sec:xid} we describe minor corrections to the official SDSS DR6 dataset,
and cross-identify the sources in the two releases.
\S~\ref{sec:def} discusses the sample selections, and we conclude
in \S~\ref{sec:sum}.
Throughout this paper, we write column names and other database
entities using typewriter fonts.

\begin{deluxetable*}{l c c c c c c}[tbh]
\tablecolumns{7} \tabletypesize{\footnotesize} \tablewidth{0pt}
\tablecaption{\small GALEX GR3 Quick Facts\label{GR3_desc}}
\tablehead{\colhead{} & \colhead{AIS} & \colhead{CAI} & \colhead{DIS} & \colhead{GII} & \colhead{MIS} & \colhead{NGS}}
\startdata
Survey area\tablenotemark{\dag} & 13,565.9 & 21.9 & 112.8 & 317.5 &
880.0 & 303.9 \\[0.1cm]
\# of fields & 15,721 & 20 & 122 & 288 &  1,017 & 296\\[0.1cm]
\# of FUV fields  & 15,721 & 16  & 98 & 275 & 1,013 & 289\\[0.1cm]
\# of NUV fields & 15,721 & 20 & 122 & 283 & 1,017 & 294\\[0.1cm]
Mean \texttt{fexptime} & 111.4 & 1,445.8  & 21,209.9 & 2,392.5 & 1,779.9 & 2,232.3 \\[0.1cm]
Mean \texttt{nexptime} & 111.5 & 2,206.6 & 26,387.0 & 3,331.6  & 2,004.2 & 2,595.0\\[0.1cm]
\# of detections & 85,358,979    & 242,526 & 2,971,137 &4,224,149  & 13,586,221 & 3,853,946 \\[0.1cm]
\# of primaries & 54,874,742  & \nodata & \nodata &\nodata  & 9,083,680 & \nodata \\[0.1cm]
\enddata
\tablenotetext{\dag}{MIS and AIS areas are given in units of square
degrees for primary footprint assuming the nominal radius of 36\arcmin{}.
Areas of other surveys assume same radius and quote unique coverage. }
\end{deluxetable*}

\section{The GALEX Catalogs}
\label{sec:galex}

The GALEX satellite observes the sky using microchannel plates in two
ultraviolet passbands: the Far-UV (FUV) centered at
$\lambda_{\rm{eff}}= 1539$\AA{} and the Near-UV (NUV) at
$\lambda_{\rm{eff}} = 2316$\AA{}.
Although the satellite also takes grism spectra, in this paper we are
only concerned with the properties of the photometric observations.

GALEX is in fact many surveys in one.
The  All-sky Imaging Survey (AIS) and the Medium Imaging Survey
(MIS) aim to systematically map the UV universe at different depths.
The Deep Imaging Survey (DIS), the Nearby Galaxy Survey (NGS), and the
Guest Investigators Survey (GII) target specific areas for various
dedicated science projects.
In addition to the above, there exists also an additional Calibration
Survey (CAI); see Table~\ref{GR3_desc} for a concise overview of the
various observation programs.

Throughout this paper, we study the properties of the catalogs in
the $3^{\rm{rd}}$ Data Release \citep[GR3;][]{morrissey07}, and use
magnitudes corrected for the Galactic extinction using the
\citet{Schlegel_1998} dust maps, with the formul{\ae} below following
\citet{Wyder_2007}.
\begin{eqnarray}
F_{\rm{corr}} & = & F - \left[8.24\,E(B\!-\!V)\right] \\
N_{\rm{corr}} & = & N - \left[8.2\,E(B\!-\!V) - 0.67\,E^2(B\!-\!V)\right]
\end{eqnarray}
Furthermore, we mainly focus on the MIS and AIS catalogs,
and their statistics.

\subsection{Primary Resolution}
\label{sec:primary}

The fields of a given survey overlap, hence certain sources are
observed many times. In the scientific analyses one would like to work
with clean catalogs that list every source only once. If multiple
detections of the same sources contaminate the dataset, the results
would be biased, and the measurements useless. While it might be
tempting to resolve this issue by selecting the best quality
observations, e.g., maximizing signal-to-noise ratio, this a posteriori selection
would create a statistical bias in the overlapping parts. Thus the preferred way
to do the primary--secondary assignment is based on prior from the survey's geometry.
Only the resolution based on
the geometry guarantees that we can maintain a good understanding of the
effective depth that often varies on the sky.

The statistical studies we are most concerned with typically rely on
the AIS and MIS datasets that both follow a common observational
strategy; namely, the field centers are targeting pre-defined
positions on the sky that we call the SkyGrid. In other words, every
given MIS or AIS field was positioned such that its center aims at one
of the pre-defined locations on the sky. This icosahedron-based tiling
algorithm is discussed in \citet{morrissey07}. The SkyGrid consists of
47,612 points on the celestial sphere that in turn define disjoint
cells on the sky in the following sense: any point on the celestial
sphere can be assigned to a single SkyGrid position by picking the
closest of them all. The contiguous region where all points belong to
the same grid point is a SkyGrid cell. These cells are all disjoint
and their union is the entire sky.
This process of subdivision is called the \citet{voronoi}
tessellation, and these cells defined by the SkyGrid centers are a
particular tessellation of the sky.

\begin{figure*}
\epsscale{1.2}
\plotone{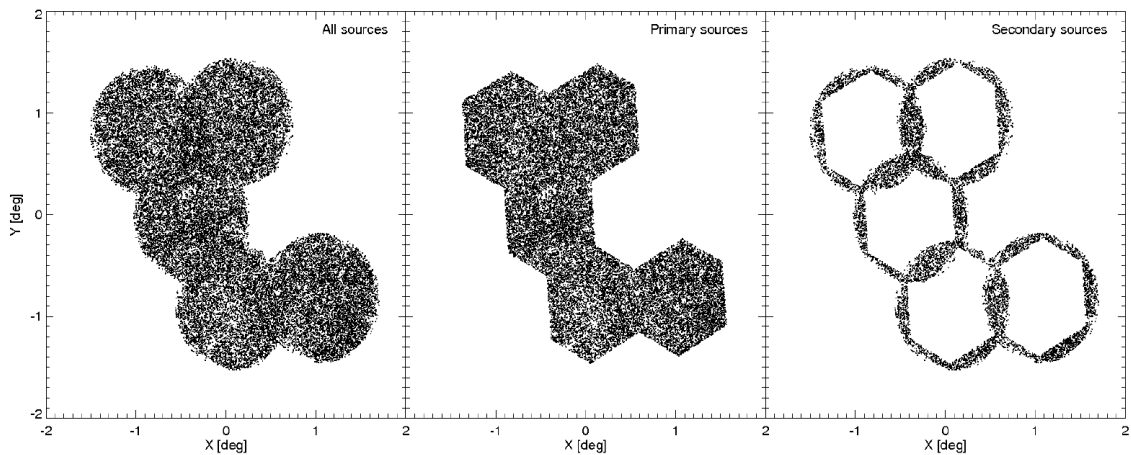}
\caption{Sources in a few random MIS fields shown in a tangent plane projection.
The left panel shows all observed sources that are split up into primaries and secondaries
seen in the middle and right panels, respectively.
See text for a detailed discussion on the advantages of primaries.}
\label{fig:primary}
\end{figure*}

If any given point is inside one and only one of the SkyGrid cells
then we can use this information to resolve the problem of multiple
detections of sources. The idea is that when a source is seen in
multiple fields, one is to take the detection that is inside the
SkyGrid cell that belongs to its own field. These objects we call
{\em{primaries}} and all the other detections are the
{\em{secondaries}}.  Of course, every secondary is going to be a
primary in some other neighboring field---assuming the field is
observed and actually covers that region. This also means that every
field has a primary region and a secondary part, and their shapes
can be defined mathematically, see later in
\S~\ref{sec:footprint}.  It also happens that primaries are
typically toward the center of the field, where the astrometric and
photometric measurements are most reliable, and secondaries tend to be
on the outskirts of the field of view. This scheme has a number of
advantages compared to other strategies. For example, it is
straightforward to add new observations to existing catalogs without
changing the existing data and their primary vs.\ secondary designations.
Figure~\ref{fig:primary} illustrates the positions of the primary and
secondary sources in a few MIS GALEX fields. The hexagon-like shapes
are typical for the primary regions. The important thing to note is
that by separating the objects into primaries and secondaries, we
create two sets of detections, where the former (middle panel)
consists of the better quality observations and has a well understood
selection function, and the latter (right panel) contains all the
rest. Despite the complicated nature of the secondaries, they are
still invaluable for various science studies, where multiple
observations of the same objects are desired, e.g., variability
analyses.

\begin{deluxetable}{l r l }
\tablewidth{0pt}
\tablecolumns{3} \tabletypesize{\footnotesize} \tablewidth{0pt}
\tablecaption{\small GALEX Mask Flags\label{table_mask}}
\tablehead{\colhead{Bit} & \colhead{Value} & \colhead{Description}}
\startdata
0 & 1 \phantom  & Detector bevel edge reflection \\[0.1cm]
1 & 2 \phantom  & Detector window reflection     \\[0.1cm]
2 & 4 \phantom  & Dichroic reflection            \\[0.1cm]
5 & 32 \phantom  & Detector rim proximity at $R=0.59^{\circ}$
\enddata
\end{deluxetable}

There still remains the issue of duplicated fields; grid cells that
were targeted multiple times. Although these are very rare and are
not intentionally part of the data releases, both AIS and MIS contain
examples by accident. Our solution is to include only the longest
exposure-time field per cell and reflect this in the primary flag of all sources in the
other fields; see details in \S~\ref{sec:workprim}.

Primaries make up roughly 65\% of the AIS or MIS detections. All
numbers quoted regarding these two surveys hereafter refer to primary
sources, unless stated otherwise. We do not define the primary
resolution for other surveys, as they do not follow the same (or any)
systematic pattern on the sky.

\subsection{Censoring Known Artifacts: Masks}
\label{sec:mask}

The GALEX photometric pipeline not only produces the calibrated images
but also builds models for potential artifacts based on the pointing
of the telescope and other variables. These models are represented as
{\em{flagmap}} images whose values describe the potential issue at
that particular pixel position.
Table~\ref{table_mask} describes the four separate layers that the
different flag values indicate.

We use these maps to censor our datasets, and mask out the artifacts.
Our approach is again a geometrical one in order to maintain a clear
understanding of the angular selection function. 
We use the
\citet{hk76} 
percolation algorithm to identify the clusters of
contiguous pixels in the flagged regions, and
these islands are then grown by a pixel to ensure that the final
representation of the boundary
encloses the original shape.
We derive the convex hull of every island separately in pixel coordinates.
These are polygons that accurately capture the outlines of these problematic shapes.
With the
world-coordinate transformations of the images \citep[WCS;][]{wcs1,wcs2} in
hand, we then convert the enclosing pixel polygon into a spherical
region that describes the censored area on the celestial sphere.
We call these {\em{}masks}.

The artifact flags are also stored for all detections and for both bands;
the properties are called \texttt{fuv\_artifact} and
\texttt{nuv\_artifact} consistently in all data release products
including the FITS files and the database servers.

\subsection{Sky Coverage}
\label{sec:footprint}

It is absolutely vital to have a precise geometric representation of
the surveys' sky coverage, and our primary designation is a
great first step in establishing the formal description of the exact
footprint. The hexagon-like cells of the primaries are simple
spherical polygons described by a half dozen (R.A., Dec) points. The
union of all these polygons is a good first approximation of the sky
coverage, but not good enough.

\begin{figure*}
\epsscale{0.85}
\begin{center}
\plottwo{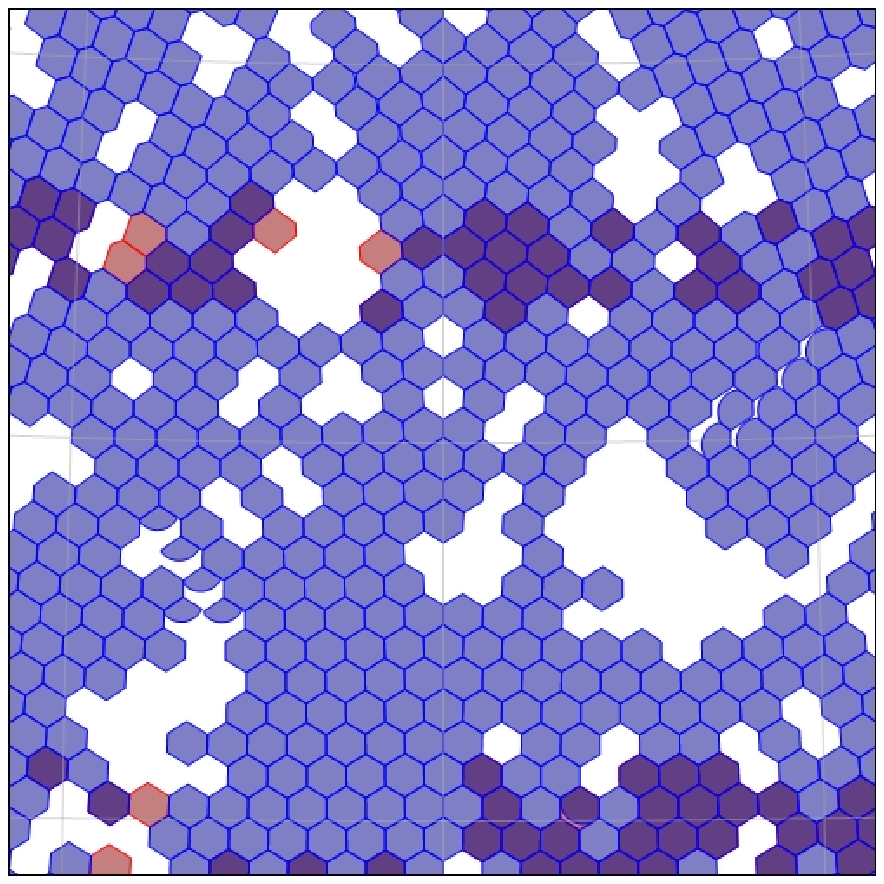}{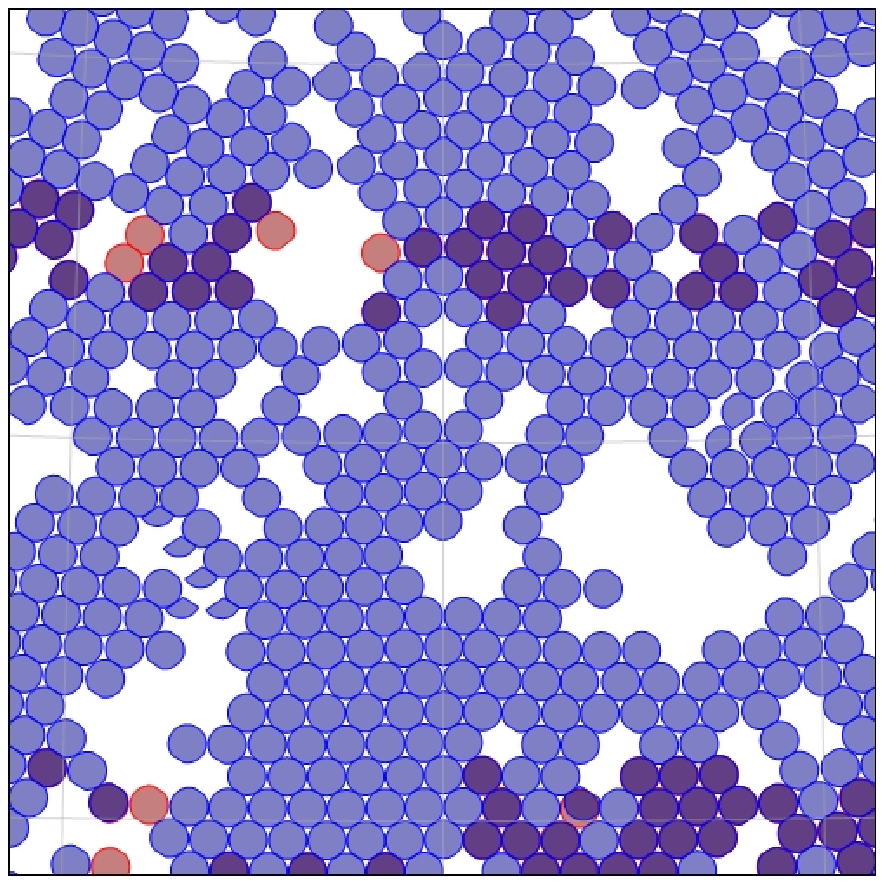}
\end{center}
\caption{These closeups of MIS and AIS primary cells (shown in red and blue,
respectively) illustrate the level of details in the sky coverage representation
in a stereographic projection at a randomly chosen position on the sky;
$(\alpha,\delta)=(0^{\circ},10^{\circ})$.
On the left, the fields are limited by the nominal 36\arcmin{} radius
circle around the field center; on the right, the limit is a more
conservative 30\arcmin{} radius. Note some of the cells have partial
coverage even in the left panel, where the pointing of the field was
considerably off from the targeted position.}
\label{fig:zoom}
\end{figure*}

So far, we have not used anywhere the information where the telescope
was actually pointing when the field was observed, only what cell the
survey was targeting with that exposure. In an ideal world these two
would be the same, but in practice they are not and sometimes the
offsets are quite significant. We need to include an additional
constraint on top of the primary cell definition that describes the
field of view. For GALEX, this is a circle around the true field
center (\texttt{avaspra}, \texttt{avaspdec}) with a nominal field
radius of 36\arcmin{} or could be a more conservative 30\arcmin{}
threshold. In Figure~\ref{fig:zoom} we show the MIS and AIS (primary)
coverage in a randomly chosen small patch of the sky using a
stereographic projection centered on
$(\alpha,\delta)=(0^{\circ},10^{\circ})$. The footprint in the left
(right) panel assumes a 36\arcmin{} (30\arcmin) radius field of view.
Note that some of the SkyGrid cells have partial coverage even with
the 36\arcmin{} radius, where the pointing of the field was
considerably off from the targeted position.

\begin{figure*}
\epsscale{0.85}
\plotone{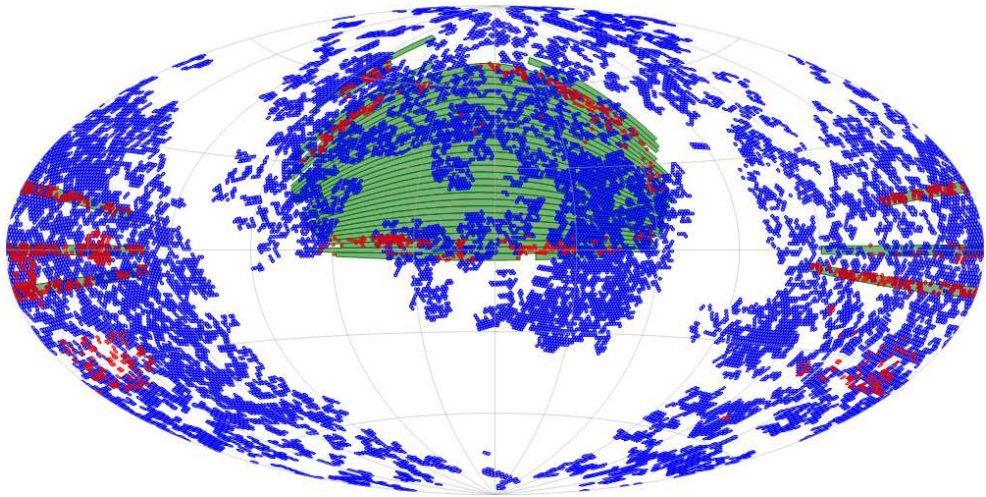}
\caption{Footprints of the SDSS (green), GALEX AIS (blue) and
MIS (red) imaging surveys, using an Aitoff projection of the equatorial J2000 system
centered on the usual $(\alpha,\delta) = (180^{\circ},0^{\circ})$ coordinates.}
\label{fig:footprints}
\end{figure*}

We use a generic mathematical framework \citep{szalay05} and a
light-weight but high-performance spherical geometry library to express all
these in a uniform way by using the halfspace, convex, region concepts
detailed in \citet{budavari_nvobook}. For every field, we intersect
the spherical polygon of the primary cell with the small circle
constraint of the field of view, and take a union of all fields to
arrive at the exact footprint description.
The GALEX sky coverage is available online on the US National Virtual
Observatory's \citep[NVO;][]{nvobook} Footprint Service%
\footnote{Visit the NVO Footprint Service at
\url{http://www.voservices.net/footprint/}} \citep{budavari_adass_tucson}.
A screenshot of the web site is shown in Figure~\ref{fig:footprints},
where the GR3 AIS and MIS footprints are compared to the SDSS DR6
photometric sky coverage in an Aitoff projection.

On top of the footprint, one needs to consider the masks; see
\S~\ref{sec:mask}.
The masks are defined per field but may very well extend beyond the
limits of the primary cell. The secondary parts of the masks need not
be applied on the primary catalog, so one needs to intersect the masks
with the primary cells of their corresponding fields and use that to
censor the sources.

The above mathematical representation of the footprint and the masks
not only enables fast filtering of random and simulated source
catalogs but also provides an exact (analytic) calculation of the area for a given polygon,
which is important for essentially all statistical studies,
e.g., the normalization of luminosity functions.

\subsection{Merged Catalogs}
\label{sec:bandmerge}

The GALEX photometric pipeline processes the FUV and NUV images
separately. Having run a custom, modified version of the
Source-Extractor \citep[a.k.a.\ SExtractor;][]{sextractor} on both bands
separately, the pipeline then attempts to merge the detections into
sources that are stored in the MCAT FITS files.

Naturally the sources are not exactly at the same position in the two
bands, hence the pipeline relies on the distance between the FUV and
NUV detections and implements a hard cutoff at three arcseconds of
separation, i.e., detections that are closer than 3'' are merged
into a single source \citep[if they meet the eligibility criteria in][]{morrissey07},
otherwise left alone as single-band detections.

For the merged objects, all FUV and NUV measurements are propagated
along with the measure separation. Since the NUV detections are
typically much higher signal-to-noise ratio than the FUV, where the
precision is often limited by the small number of observed photons,
the merged object carry the position of the NUV detections by default.
While this a sensible choice, one needs to be aware of the fact when
pinning down geometrical constraints.

\section{Cross-Identification of Sources}
\label{sec:xid}

\begin{deluxetable}{l c r r}
\tablecolumns{4} \tabletypesize{\footnotesize} \tablewidth{0pt}
\tablecaption{\small  Number of matched sources by SDSS types\label{galexxdr6_types}}
\tablehead{\colhead{Classifier} & \colhead{Type} & \colhead{AIS} &  \colhead{MIS} }
\startdata
Photo & galaxy & 4,542,255   & 2,651,799  \\[0.1cm]
      & star & 2,973,395 &  1,059,731 \\[0.1cm]
\hline
Spectro & galaxy  & 132,968 &  37,060  \\[0.1cm]
 & star & 20,930  &   4,284   \\[0.1cm]
 & quasar  & 30,860  &  6,419   \\[0.1cm]
 & high-$z$ quasar  & 524  &  126   \\[0.1cm]
\enddata
\tablecomments{Numbers are given for one-to-one matches of
primary sources in both GALEX and SDSS.}
\end{deluxetable}

The cross-identification of sources in various catalogs is fairly
complicated in general. One needs a good understanding of the
astrometry of the observations involved and their sky coverage.
With the footprint descriptions in hand, one can decide whether a
missing counterpart is truly a dropout or if that part of the sky was
just simply not covered by the other observation.
Based on the astrometric precisions one can assign a probability to a
set of detections in separate catalogs that determines whether they
belong to the same object. \citet{budavari_szalay} introduced a
Bayesian approach to the matching problem, and showed that for the
usual astrometric model, the spherical normal distribution
\citep{fisher53}, the problem can be analytically integrated.
The observational evidence for a set of sources being the same object
is calculated as a function of their separations and, of course, the
astrometric precisions.

\subsection{SDSS Versus GALEX}
\label{sec:xmatch}

The SDSS data releases are accompanied by concise descriptions
on the Project's web site%
\footnote{Visit the SDSS Project at \url{http://www.sdss.org/}}
as well as
refereed science papers, and the latest 6th Data Release we
focus on here is no exception \citep[DR6;][]{dr6}.
We applied only a few modifications to correct for some
minor flaws in DR6 involving the primary resolution of a handful of
sources, i.e., duplicate primaries in Stripe 36 and extra fields outside
the primary footprint in Stripes 38 and 44,
and incorrect \texttt{SkyVersion} in some of the mask identifiers. These issues
were discovered after the catalogs went
public, and while the upcoming new catalogs will not have these
issues, the official DR6 dataset was frozen with the release.

When the probabilistic cross-identification methodology is applied to
the SDSS and GALEX cases with the nominal $\sigma=0.1\arcsec$
\citep{pier} and $0.5\arcsec$ \citep{morrissey07} accuracies,
respectively, we find the separation limits as a function of the
probability threshold. For matching GALEX sources to SDSS, the Bayes
factor becomes unity at less than 4\arcsec. This does not mean that
one has to accept all matches within 4\arcsec, but rather it provides
a safe search radius for selecting candidates. This limit is
approximately 5\arcsec{} when matching GALEX to itself. Also, we find
that for MIS the probability limit of 50\% is at around 3\arcsec.

\begin{figure}
\epsscale{1.2}
\plotone{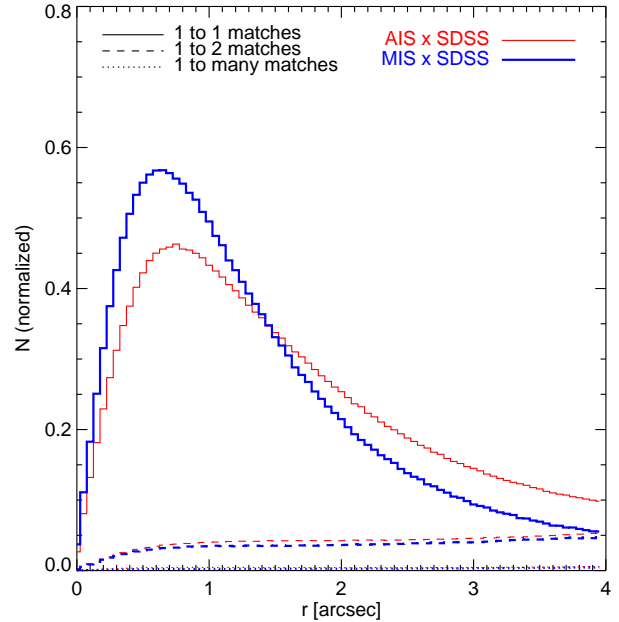}
\caption{Distributions of pairwise distances for GALEX-SDSS
primary matches: one-to-one matches (solid lines), one-to-two matches (dashed
line) and one-to-many matches (dotted lines). These distribution are
plotted for the AIS (thin red) and MIS (thick blue) surveys.}
\label{fig:pairwise}
\end{figure}

\begin{deluxetable}{l c c c c c c}
\tablecolumns{7} \tabletypesize{\footnotesize} \tablewidth{0pt}
\tablecaption{\small GALEX-SDSS Quick Facts\label{galexxdr6_stat}}
\tablehead{\colhead{} & \colhead{AIS} & \colhead{CAI} & \colhead{DIS} & \colhead{GII} & \colhead{MIS} & \colhead{NGS}}
\startdata
Area of Intersection\tablenotemark{\ddag}  & 3645.8 & 10.0 & 70.6 & 132.6 & 598.1 & 104.6 \\
Number of GALEX Fields                                & 4748 & 9 & 79 & 130 & 786 & 105 \\
Number of Matches & 9,463,978  & \nodata & \nodata & \nodata & 4,525,588 &\nodata \\
\enddata
\tablenotetext{\ddag}{In units of square degrees} 
\end{deluxetable}

Motivated by the analytic results, we found the candidate sources in
the GALEX merged catalogs and the SDSS dataset by accepting the
separation threshold of 4\arcsec{}. The matching is done entirely
inside the relational database engine (SQL Server) that holds the
GALEX and SDSS science archives using advanced indexing schemes to
make it not only feasible but fast \citep{gray06, nieto_phd}.
The generated slim table that connects the sources is essentially a
many-to-many mapping, where most GALEX sources have only one SDSS
counterpart but other combinations also occur frequently.
In Table~\ref{galexxdr6_types} we list the number of one-to-one
GALEX-SDSS matches for spectroscopically confirmed stars, galaxies and
quasars (\texttt{SpecClass}), as well as broken down by the SDSS
star/galaxy separation (\texttt{Type}), which is based on profile
fitting.
The distributions of pairwise distances are shown
Figure~\ref{fig:pairwise} for the AIS and MIS surveys.  The
distribution is slightly broader for the AIS matches compared to the
MIS case, which is due to the lower signal-to-noise ratio that yields
lower accuracy, cf.\ measurements in \citet{morrissey07}.

Table~\ref{galexxdr6_stat} summarizes the basic properties of the associations.
We enumerate the number of fields in the various GALEX surveys that overlap with SDSS
and the number of sources within.
In addition we also list the analytic area calculations of the interesections.

\subsection{Visual Inspection}
\label{sec:cutout}

To visually check the GALEX-SDSS cross-identifications, and to compare
UV to optical imaging of the sources in general, we developed a new
online tool that displays the false-color images of the two surveys
side by side. Our Image Cutout%
\footnote{Visit the GALEX-SDSS Cutout at \url{http://voservices.net/galex/cutout/}}
is available for the public.
The web application programmatically accesses the SDSS cutout service on the SkyServer%
\footnote{Visit the SDSS SkyServer at \url{http://skyserver.sdss.org/}}
and displays that image without modification next to the GALEX image.
The GALEX mosaics combine all information in the FUV and NUV images to
provide the most detail. The brightness of the pixels is a function of the
signal-to-noise ratios of the two UV bands added in quadrature
\citep[$\chi^2$ image;][]{chisq}. The mapping is the asinh() function first
introduced for the SDSS magnitudes \citep{lupton_asinh} that behaves
like the logarithm in the classic magnitudes for bright sources but
becomes linear for small fluxes.
The color in each pixel encodes the ratio of the FUV to the NUV
fluxes. We build the cumulative distribution of this ratio, suitably
normalized in order to span the entire color palette. This
distribution is then fitted by an analytical function using
an atan() function. The Hue value in the HSV color space
is then mapped from this fit the given FUV to NUV
ratio.

\begin{deluxetable}{l r r r }
\tablecolumns{4} \tabletypesize{\footnotesize} \tablewidth{0pt}
\tablecaption{\small Number of AIS matches by association\label{tab:matrix_ais}}
\tablehead{
           \colhead{GALEX \raisebox{1.5ex}{SDSS}} &
           \colhead{\raisebox{1.5ex}{1}} &
           \colhead{\raisebox{1.5ex}{2}} &
           \colhead{\raisebox{1.5ex}{Many}}}
\startdata
1 & 7,522,205 & 1,213,972 & 125,423 \\[0.1cm]
2 & 507,656 & 83,657 & 7,953 \\[0.1cm]
Many & 2,598 & 455 & 59 \\[0.1cm]
\enddata
\end{deluxetable}

\begin{deluxetable}{l r r r }
\tablecolumns{4} \tabletypesize{\footnotesize} \tablewidth{0pt}
\tablecaption{\small Number of MIS matches by association\label{tab:matrix_mis}}
\tablehead{
           \colhead{GALEX \raisebox{1.5ex}{SDSS}} &
           \colhead{\raisebox{1.5ex}{1}} &
           \colhead{\raisebox{1.5ex}{2}} &
           \colhead{\raisebox{1.5ex}{Many}}}
\startdata
1 & 3,712,815 & 525,924 &  40,497 \\[0.1cm]
2 &  215,509 &  28,647 & 1,962 \\[0.1cm]
Many & 200 & 29 & 5 \\[0.1cm]
\enddata
\end{deluxetable}

The number of one-to-two (one GALEX to two SDSS) matches increases slowly with
distance, which is the expected trend for random associations. The one-to-many
cases may occur for various reasons: at bright magnitudes,
because of shredding in SDSS, and at faint magnitudes due to objects
blending in GALEX. 
The fraction of one-to-two (one-to-many) matches
depends on the UV magnitude: it is roughly constant at 20\% (5\%) up to
FUV=19 and decreases to 10\% (0\%) at FUV=24. For NUV-selected
sources, the fraction is fairly constant at 15\% (5\%) up to NUV=18
and then decreases to 10\% (0\%) at NUV=24.
In Tables~\ref{tab:matrix_ais} and \ref{tab:matrix_mis}, we enumerate
the frequency of the various cases.

The other large component in the contingency matrices is the two-to-one
matches, where there are two GALEX detections for every SDSS source.
Our visual inspection of a large number of these cases unravels a
peculiar fact: most of these associations, roughly 85\%, have GALEX
sources, where one out of the two is detected in the NUV only and the
other in only the FUV. This suggests the possibility that all these
detections are, in fact, from the same source but they were not merged in
the pipeline because they were either not eligible to be merged by not
meeting the signal-to-noise ratio limit, or their distances are greater
than the 3\arcsec{} limit.

\begin{figure}
\epsscale{1.2}
\plotone{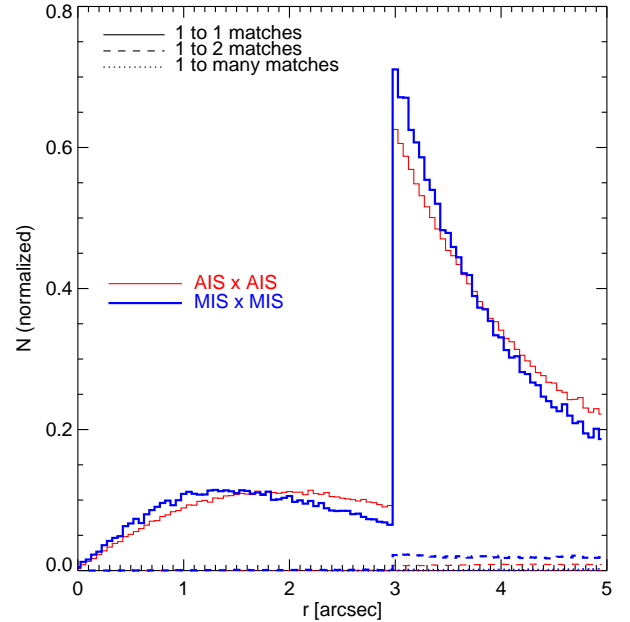}
\caption{Distributions of pairwise distances for GALEX-GALEX
primary matches: one-to-one matches (solid lines), one-to-two matches (dashed
lines) and one-to-many matches (dotted lines). These distribution are
plotted for the AIS-AIS (thin red) and MIS-MIS matches (thick blue). The break at
3\arcsec{} is due to the fact that FUV and NUV sources at smaller separations
are already merged by the pipeline prior to our cross-identification.}
\label{fig:pairself}
\end{figure}

\subsection{Revisiting GALEX Band-Merging}
\label{sec:revise}

We performed the cross identification of the GALEX sources to
themselves (excluding the identities) with a search radius of
5\arcsec{} using the nominal positions, (\texttt{ra},
\texttt{dec}). This limit is again motivated by the Bayesian analysis,
and should provide a safe margin to elect all candidates. The pairwise
distance distributions of primaries are shown in
Figure~\ref{fig:pairself}.  We see a sharp break at
3\arcsec{}, the limit of the official band merging.  At smaller
separations than the break, there can be sources that were very close
to the SNR threshold: although they made the detection limit of
SExtractor, they were not eligible to be merged. These two thresholds
are both roughly at around $2\sigma$ but not exactly the same in the
two applications. Alternatively these could be cases where one of the
sources in the pair are already merged with some another source. 
We find that the latter is not the case.  In fact, the whole histogram is
dominated entirely by NUV/FUV only detections. Below the break, the
FUV fluxes are essentially in the noise. Above the 3\arcsec{}
separation, it is again FUV/NUV only detections of any signal-to-noise
ratio. Our statistics are also in accord with the visual inspection
described in the previous section.

\begin{figure*}
\epsscale{1}
\plotone{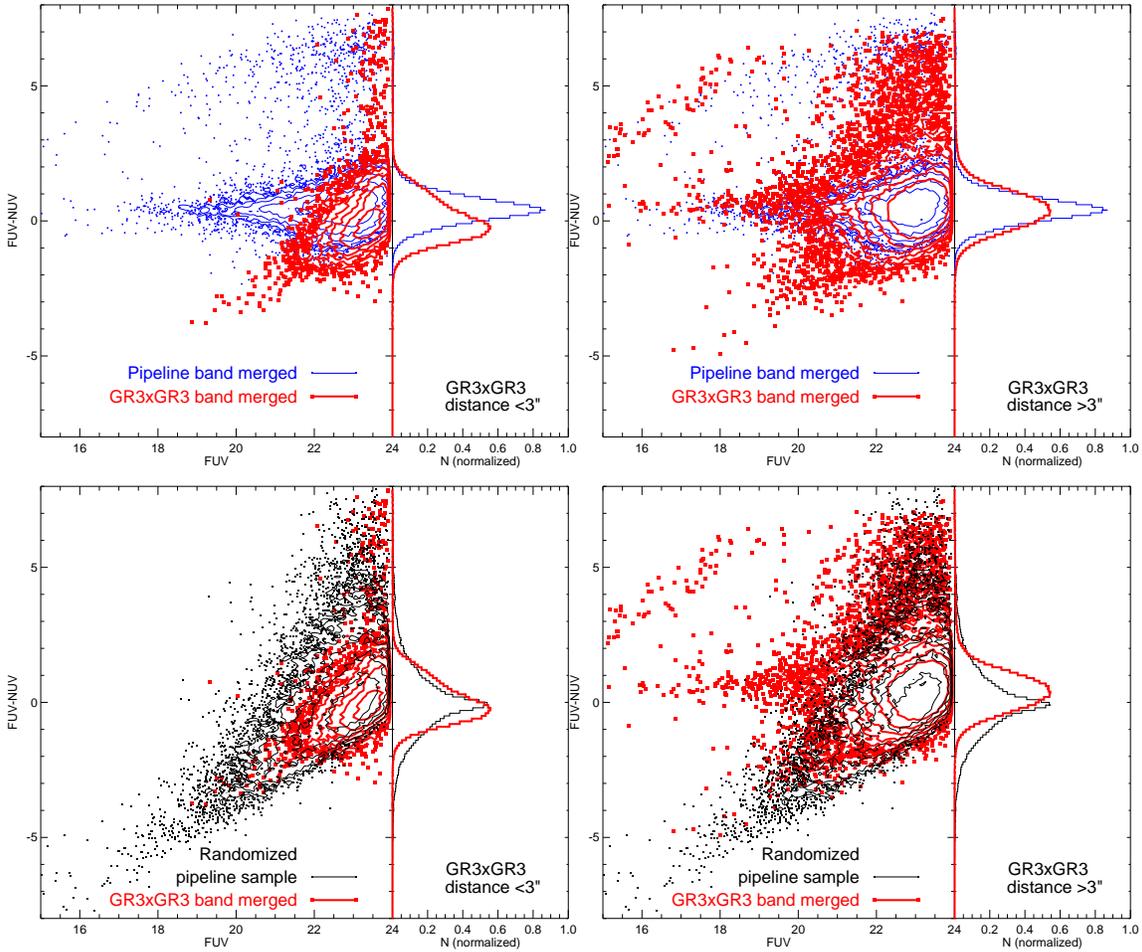}
\caption{Color-magnitude diagrams for MIS pipeline-merged sources (thin blue in top panels),
random FUV-NUV associations for reference (thin black in bottom panels) and merged sources
from the GR3-GR3 cross-identifications (thick red). The left and right panels show
the results for pairs with smaller and larger than 3\arcsec{} separations.
The insets on the right
in all panels show the normalized histograms of the corresponding
samples.}
\label{fig:colors}
\end{figure*}

\begin{figure}
\epsscale{1.15}
\plotone{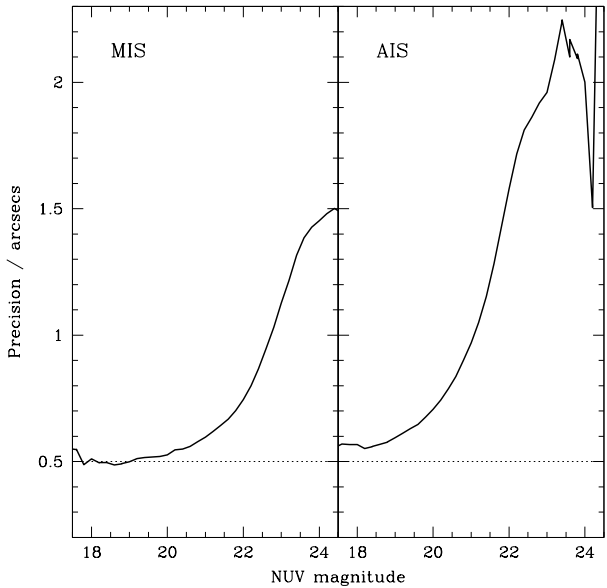}
\caption{Precision of the GR3 astrometry in the MIS (left) and AIS (right)
as a function of the apparent NUV magnitude.
The precision of GALEX detections is measured by the median angular
separations from the corresponding
SDSS DR6 point sources. Only one-to-one match primaries are considered.}
\label{fig:astrometry}
\end{figure}

The census of the merged FUV-NUV pairs inside and beyond the official
3\arcsec{} separation limit can only be done systematically in a
statistical way due to their large numbers.
Our approach is to assume that the associations are correct and study
their properties, namely the UV color-magnitude diagram, to look for
inconsistencies.
We divide the sample into two subsets based on the separations
(greater/less than 3\arcsec) as the two samples are expected to have
different properties, and compare their distributions to a couple of
reference sets. The first is the list of the merged sources produced
by the GALEX pipeline, and the other is an artificial dataset, where
one randomly shuffles the NUV-FUV associations. These two
distributions are naturally quite different.
If the distribution of the matched detections follows the trends
seen in the pipeline data, we can be confident that they are typically
good associations, and if they are more similar to the randomized
dataset, they are mostly noise.
Figure~\ref{fig:colors} shows the results of these comparisons in
four panels. Each panel contains an inset on the left with the
color-magnitude diagram and the normalized color histogram on the
right. The left panels illustrate the small separation subsample, and
the right panels the pairs with large distances. The top and bottom
two panels, compare these distributions to the different reference
sets: pipeline on top, the randoms below.
We see that the associations with smaller than 3\arcsec{} separations
look more like the randoms, although the asymmetric shape of the color
histogram suggests that there are real objects in the mixture, as
well. Knowing that these are very low signal-to-noise detections, this
result is what one would expect.
On the right, we see that on the other side of the 3\arcsec{} break
the sources look very much like the pipeline colors. Since we elected
the 5\arcsec{} matching radius to be a safe cut, it seems odd at first
that, in fact, most of these associations with these larger
separations have meaningful UV colors. The implication of these
results is that the astrometric precision is less accurate than the
nominal $\sigma=0.5\arcsec$. We find independently that the limitation
of the positional accuracy is set by the low photon counts in the FUV
detectors, which is being studied and will be better assessed for the
upcoming data releases. In Figure~\ref{fig:astrometry} we look at the
positional differences between GR3 and DR6 primaries, using only one-to-one matches,
to quantify the accuracy as a function of the NUV magnitude. For this
measurement we selected a clean subsample of SDSS point sources. In accord
with our expectations, we see that the accuracy gets worse with the magnitude
and it is often higher than 0.5\arcsec{}.

Since the colors of the associations at these larger separations out
to 5\arcsec{} statistically prove to be physical, one can safely
include large fraction of these extra merged sources that are quite significant in number.
The increase in the size of the merged catalog is up to 12\%
for MIS, and
even higher, due to the lower signal-to-noise ratios, for AIS up to 26\%.

\begin{figure*}
\epsscale{1}
\plotone{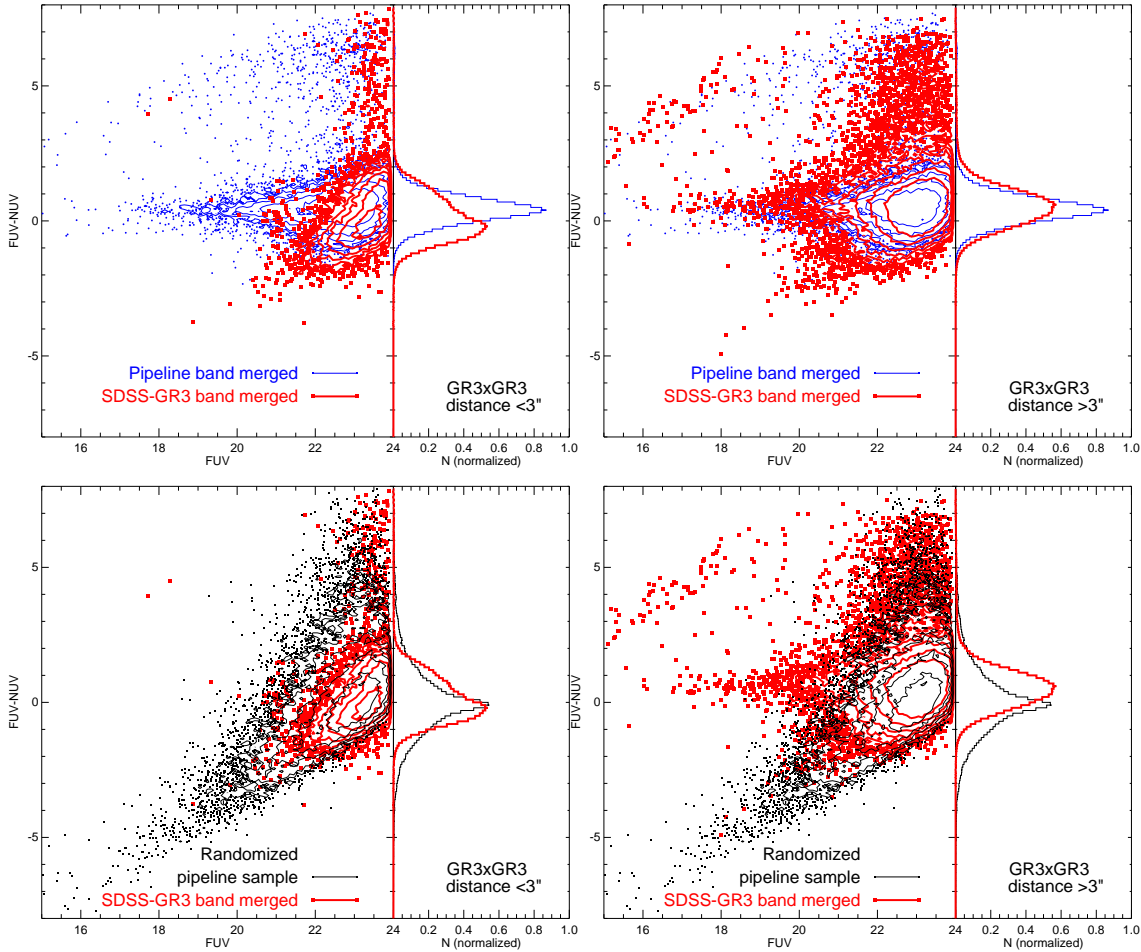}
\caption{Same as Figure~\ref{fig:colors} for GALEX associations that share a common SDSS counterpart.}
\label{fig:colors2}
\end{figure*}

In Figure~\ref{fig:colors2} we plot the same four panels for GALEX
associations that share a common SDSS object. While the basic
characteristics of the figures are essentially identical to the
previous, one sees a difference in the wings of the color
distributions, which are slightly tilted away from the artificial
random colors. Using this extra constraint promises to improve the
detection limit when desired.

We estimate the fraction of extra merged sources with separations
larger than 3\arcsec{} that have similar properties than pipeline
sources within the UV color magnitude-diagram. To that aim, we assume
that the distribution of the extra merged sources is a linear
combination of pipeline and random data. The overall fraction of extra
merged sources similar to pipeline data is $\sim 0.6$ if we consider
only GALEX associations. Using the associations that share a common
SDSS object yields a larger fraction: $\sim 0.65$. The results depend
on magnitude, as most of the objects with pipeline-like colors are
fairly faint; for objects with 20$<$FUV$<$22, the fraction is 0.65 (0.7
with SDSS constraint); for fainter objects, 22$<$FUV$<$24, the fraction
increases to 0.7 (0.8 with SDSS constraint).

\section{Defining the Catalogs}
\label{sec:def}

Understanding the GALEX reduction pipeline and the properties of the
extracted source catalogs is crucial for working with the data. We use the
GALEX dataset as published by the Multimission Archive at the Space
Telescope Science Institute (MAST)%
\footnote{Visit the GALEX database at \url{http://galex.stsci.edu}}
and augment the database with auxiliary tables and properties. 
The additions include tables to describe the geometry of the SkyGrid in
form of the database table \texttt{SkyGridV2} and the information to link
the cells to the GALEX fields via the \texttt{GridID} column in the table
\texttt{PhotoExtract}. Some of the changes had been propagated back to
the official MAST site but the full update is expected in early 2009
along with the new GALEX and SDSS releases (GR5 and DR7) for which this paper also
serves as a guidebook and documentation.

Due to the large overlap between the GALEX and SDSS sky coverage,
it is best to keep both datasets in the same logical framework to perform
meaningful selections in reasonable times. 
MAST is expected to start serving the associations early next year.
To accommodate the typical science queries, we will provide early 
access to the GR3 and DR6 cross-match catalog via the CasJobs site%
\footnote{Visit the JHU CasJobs site at \url{http://skyservice.pha.jhu.edu/casjobs}}
hosted at The Johns Hopkins University. The site utilizes technologies
developed for federating archives within the National Virtual
Observatory (NVO).

\subsection{Working with Primaries} \label{sec:workprim}

As discussed in \S~\ref{sec:primary}, the use of primaries is very
advantageous since this collection includes every source only once. The
complicated geometrical selection criteria based on the SkyGrid cells
should not discourage their usage. Every GALEX source in the database
has a flag called \texttt{Mode} that provides the result of the
primary resolution. The value of this property is set to 1 for all
primaries and to 2 for secondaries. Sources that are not in AIS or MIS
will have a value of 0. By definition, any other values signal a
problem.  Figure~\ref{fig:sql-primary} illustrates an SQL request that
retrieves MIS primaries that are seen in both FUV and NUV (merged by
the photometric pipeline) if they are within 30\arcmin{} from the
center of the field. Note the field center is not the center of the
SkyGrid cell but the position encoded in the (RA, Dec) coordinates
(\texttt{avaspra}, \texttt{avaspdec}).

A simple and useful test is to look for duplicate fields. Since the
primaries are defined field-by-field, incidental multiple exports of the same
part of the sky would again result in double-counting of certain
sources. The query in Figure~\ref{fig:sql-dups} finds the problematic
cells for AIS and MIS. In the catalog up to GR3, both contain a field
each that was exported twice and one of those fields, preferably with
the less exposure time, should be excluded. Instead of removing the
fields from the data release, we simply flip a bit of the
\texttt{Mode} to exclude these sources from our statistical samples;
the values will be 5 and 6.

While we are looking at the fields, let us quickly also determine
which are the fields that are covered by both AIS and MIS. The
subsamples within them are invaluable for cross checks and studies of
the limitations. In Figure~\ref{fig:sql-common} we show an SQL command
that compares the FUV exposure times for the resulting fields.

\subsection{Working with Associations}

We provide the cross-identification of all GALEX sources to
themselves. The links stored in the table \texttt{xSelf} not only
enable quality assurance and diagnostic tests, but also deliver the
missing merged sources at separations larger than the
3\arcsec{} limit in the GALEX pipeline out to 5\arcsec{}. The query in
Figure~\ref{fig:sql-nf} emulates the pipeline merging. Here one picks
the pairs where one is NUV only detection (\texttt{Band=1}) and the
other is FUV only (\texttt{Band=2}) and only requires the NUV
detection to be primary, as it is the accepted position for the merged
source in the pipeline.

The GALEX-SDSS DR6 associations are stored in the table
\texttt{xSdssDR6}, which again represents many-many mapping between
the sources. We also introduce tables where the sources are grouped by
GALEX and SDSS identifiers and store the number of matches for each,
see tables starting with \texttt{xGroup}. Figure~\ref{fig:sql-join}
shows an SQL query that counts the number of one-to-one matches, where
both SDSS and GALEX sources are primaries and breaks the numbers down
by the SDSS object classification.

\begin{figure*}
\begin{center}
\begin{minipage}{13cm}
\begin{verbatim}
SELECT ObjID, FUV_Mag, NUV_Mag
FROM PhotoObjAll o
    JOIN PhotoExtract e ON e.PhotoExtractID = o.PhotoExtractID
WHERE e.MpsType = 'MIS'
    AND o.Mode = 1 -- i.e., primary
    AND o.Band = 3 -- both bands (1:NUV, 2:FUV)
    AND o.FoV_Radius < 0.5
\end{verbatim}
\end{minipage}
\end{center}
\caption{An example query in SQL to retrieve
all GALEX MIS primaries within 30\arcmin{} of the field centers that
have both FUV and NUV measurements merged by the photometric pipeline.
The Transact-SQL keywords are typeset in capital letters; for table
and column names, we use the conventional camel case spelling.}
\label{fig:sql-primary}
\end{figure*}

With the formal description of the GALEX footprint in hand, it is
straightforward to look for dropouts, where the SDSS source is missing
in GALEX. We achieve this by first assigning every SDSS source to a
SkyGrid cell and then selecting the cells that were actually observed
in the given survey. The table \texttt{SdssDr6inMisPrimary} contains
all the SDSS sources that are in the MIS footprint along with the
distance to the actual field center. Figure~\ref{fig:sql-dropout}
illustrates the query where the crossmatch table is used to search for
SDSS galaxies close to the field center, $d<0.5^{\circ}$, without a
counterpart.

\subsection{Revised band-merging}

We now combine the above key elements of sample selection into a real-life
example of creating a catalog of primaries with detections in both bands
using the extra associations discussed above, i.e.,
the FUV-, NUV-only matches
at larger angular separations than the official pipeline cutoff.
We propagate the basic quantities of GALEX measurements along
with links to the SDSS associations, where available.
Figure~\ref{fig:sql-master} illustrates the SQL command that defines the final catalog;
the union of the three sub-queries. These are conceptually
really only two: the pipeline merged sources (\texttt{Band=3}) and
the one-to-one GALEX matches (\texttt{Band=1} to \texttt{Band=2}).
The third part is simply a result of the fact that the match table
\texttt{xSelf} is not symmetric by construct to save storage space
and is essentially identical to the
previous one except for the order of the \texttt{ObjID}s in the join criteria.

\begin{figure}[b!]
\begin{center}
\begin{minipage}{13cm}
\begin{verbatim}
SELECT MpsType, GridID, COUNT(*)
FROM PhotoExtract
WHERE MpsType IN ('AIS','MIS')
GROUP BY MpsType, GridID
HAVING COUNT(*) > 1
-- AIS   2350  2
-- MIS  21884  2
\end{verbatim}
\end{minipage}
\end{center}
\caption{Look for SkyGrid cells in AIS and MIS that were observed
multiple times and need to be resolved for a clean sample to avoid duplicates.}
\label{fig:sql-dups}
\end{figure}

\begin{figure*}
\begin{center}
\begin{minipage}{13cm}
\begin{verbatim}
SELECT mis.PhotoExtractID AS MisID, mis.FExpTime AS MisFExpTime,
       ais.PhotoExtractID AS AisID, ais.FExpTime AS AisFExpTime
FROM PhotoExtract mis
    JOIN PhotoExtract ais ON ais.GridID = mis.GridID
WHERE mis.MpsType = 'MIS' and ais.MpsType = 'AIS'
-- 809 fields
\end{verbatim}
\end{minipage}
\end{center}
\caption{Look for fields that have been observed in both AIS and MIS.}
\label{fig:sql-common}
\end{figure*}

\begin{figure*}
\begin{center}
\begin{minipage}{13cm}
\begin{verbatim}
SELECT e1.MpsType, COUNT(*) as Number
FROM xSelf x
    JOIN PhotoObjAll o1 ON o1.ObjID = x.ObjID
    JOIN PhotoObjAll o2 ON o2.ObjID = x.MatchID
    JOIN PhotoExtract e1 ON e1.PhotoExtractID = o1.PhotoExtractID
    JOIN PhotoExtract e2 ON e2.PhotoExtractID = o2.PhotoExtractID
        AND e1.PhotoExtractID = e2.PhotoExtractID
WHERE x.Distance BETWEEN 3 AND 5
    AND e1.MpsType IN ('AIS', 'MIS')
    AND (    (o1.Band = 1 AND o2.Band = 2 AND o1.Mode = 1)
          OR (o1.Band = 2 AND o2.Band = 1 AND o2.Mode = 1)
        )
GROUP BY e1.MpsType
\end{verbatim}
\end{minipage}
\end{center}
\caption{SQL query that counts the number of NUV-only primaries in AIS and MIS
that are matched to FUV-only detections in the same field with separations
between 3\arcsec{} and 5\arcsec{}.}
\label{fig:sql-nf}
\end{figure*}

\begin{figure*}
\begin{center}
\begin{minipage}{13cm}
\begin{verbatim}
SELECT x.MatchType, COUNT(*)
FROM xSdssDr6 x
    JOIN xGroupMIStoSdssDr6Primary gs ON gs.ObjID = x.ObjID
    JOIN xGroupSdssDr6toMISPrimary sg ON sg.MatchID = x.MatchID
WHERE x.Mode = 1 AND x.MatchMode = 1 -- both primaries
    AND gs.N = 1 -- only 1 SDSS match for GALEX source
    AND sg.N = 1 -- only 1 GALEX match for SDSS source
GROUP BY x.MatchType -- e.g., 3:Galaxy, 6:Star
\end{verbatim}
\end{minipage}
\end{center}
\caption{Returns the number of one-to-one matches of
GALEX and SDSS primaries broken down by the
SDSS photometric type classification.}
\label{fig:sql-join}
\end{figure*}

\begin{figure}
\begin{center}
\begin{minipage}{13cm}
\begin{verbatim}
SELECT COUNT(*)
FROM SdssDr6inMisPrimary s
    LEFT OUTER JOIN xSdssDr6 x ON x.MatchID = s.ObjID
WHERE s.Type = 3 -- i.e., galaxy
    AND s.Distance < 0.5 -- from the field center
    AND x.ObjID IS NULL -- no match
\end{verbatim}
\end{minipage}
\end{center}
\caption{Returns the number of SDSS primaries in the MIS footprint
that are closer to the field center than 30\arcmin{} but have no GALEX
counterparts, a.k.a.\ dropouts.}
\label{fig:sql-dropout}
\end{figure}

We select the most common properties sufficient for many
scientifically interesting queries but note that any other attributes are also easily
selected by joining the result of this query with the original
\texttt{PhotoObjAll} table. The columns we choose to propagate are the
following: (1) \texttt{MpsType} that takes the values of AIS and MIS,
(2) the field identifier \texttt{PhotoExtractID}, (3-4) the identifiers
of the FUV and NUV observations, which are the same for the merged objects,
(5-6) NUV position, (7-8) magnitudes, (9) angular separations in arcseconds
between the FUV and NUV detections, and (10) the identifier of the SDSS primary
associations and (11) their distances in arcseconds, where available.
The first query simply picks GALEX primaries with both bands but the anatomy of the
following is more complicated: We use the aforementioned \texttt{xSelf} table to link
NUV-only primaries to FUV-only sources within the same fields that are farther
than 3\arcsec{}. Note that because the we elect the NUV coordinates to be the position
of the merged source following the pipeline strategy, we do not require the FUV-only
source to be a primary. We also ensure that only one-to-one GALEX matches are used,
which is essentially everything. The final size of the AIS and MIS two-band catalogs grow
from 2,634,974 to 3,819,307 and  1,549,355 to 1,988,882, respectively.

%

Another approach to improve on the colors of the sources
is to use a separate set of FUV fluxes from the GALEX pipeline: The FUV flux within the
NUV aperture is also published in the catalog called {\em{fd-ncat}}, whose parameters
are also found in the \texttt{PhotoObj} table. For the DIS catalogs, where the confusion
becomes an issue, there is a new algorithm being developed to use optical and/or NUV prior
information on the positions.
The potential problem with the {\em{fd-ncat}} measurements
arises when the NUV and FUV positions are significantly different and the NUV aperture encloses
only part of the FUV object, which yields typically
smaller FUV fluxes than the actual. This could considerably affect techniques that strongly
rely on the color, such as the FUV dropout selection of the
{\em{Wiggle-Z}} project \citep{wigglez}. 

\section{Summary}
\label{sec:sum}

We presented a detailed study of the GALEX photometric catalogs. Our
discussion focused mostly the AIS and MIS datasets and their
properties. We define the primary area of the fields to resolve
duplicates in overlapping fields and the sky coverage. We also derived
masks that are stored along with the sources in spherical polygons. We
performed the cross-identification of the GALEX GR2+3 and SDSS DR6
sources and indicated issues with multiple matches that often point
toward unmerged GALEX detections that are only seen in the NUV and in
the FUV. We found that merging these detection can be safely done out
to 5\arcsec{} separations, which also implies that the nominal
$\sigma=0.5\arcsec$ astrometric precision is somewhat optimistic.  We
made the cross-match catalogs available to the general public in the
form of a SQL Server database engine and showed various examples of SQL
queries to not only access the data but to extract clean catalogs for
scientific analysis.

\begin{figure*}
\begin{center}
\begin{minipage}{13cm}
\begin{verbatim}
SELECT e.MpsType, o.PhotoExtractID,
    o.ObjID as FuvObjID, o.ObjID as NuvObjID,
    o.RA, o.Dec, o.Fuv_Mag, o.Nuv_Mag,
    60*dbo.fDistanceArcMinEq(Nuv_RA,Nuv_Dec,Fuv_RA,Fuv_Dec) as Separation,
    gs.MatchID as SdssObjID, gs.Distance as Distance
FROM PhotoObjAll o
    JOIN PhotoExtract e ON e.PhotoExtractID=o.PhotoExtractID
    LEFT OUTER JOIN xSdssDR6 gs
        ON gs.ObjID = o.ObjID AND gs.MatchMode = 1
WHERE Band = 3 AND o.Mode = 1 -- Primaries with both bands

UNION ALL

SELECT en.MpsType, n.PhotoExtractID,
    f.ObjID as FuvObjID, n.ObjID as NuvObjID,
    n.RA, n.Dec, f.fuv_mag as FUV_mag, n.nuv_mag as NUV_mag,
    gg.Distance as Separation,
    gs.MatchID as SdssObjID,
    gs.Distance as Distance
FROM xSelf gg
    JOIN PhotoObjAll f ON f.ObjID = gg.ObjID
    JOIN PhotoObjAll n ON n.ObjID = gg.MatchID
    JOIN PhotoExtract ef ON ef.PhotoExtractID = f.PhotoExtractID
    JOIN PhotoExtract en ON en.PhotoExtractID = n.PhotoExtractID
        AND ef.PhotoExtractID = en.PhotoExtractID
    JOIN xGroupAMIStoAMIS g1 ON g1.ObjID = n.ObjID
    JOIN xGroupAMIStoPrimary g2 ON g2.ObjID = f.ObjID
    LEFT OUTER JOIN xSdssDR6 gs
        ON gs.ObjID = n.ObjID AND gs.MatchMode = 1
WHERE gg.Distance BETWEEN 3 AND 5
    AND n.Band = 1 AND f.Band = 2 AND n.Mode = 1
    AND g1.N = 1 AND g2.N = 1 -- One-to-one GALEX match

UNION ALL

SELECT en.MpsType, n.PhotoExtractID,
    f.ObjID as FuvObjID, n.ObjID as NuvObjID,
    n.RA, n.Dec, f.fuv_mag as FUV_mag, n.nuv_mag as NUV_mag,
    gg.Distance as Separation,
    gs.MatchID as SdssObjID,
    gs.Distance as Distance
FROM xSelf gg
    JOIN PhotoObjAll f ON f.ObjID = gg.MatchID
    JOIN PhotoObjAll n ON n.ObjID = gg.ObjID
    JOIN PhotoExtract ef ON ef.PhotoExtractID = f.PhotoExtractID
    JOIN PhotoExtract en ON en.PhotoExtractID = n.PhotoExtractID
        AND ef.PhotoExtractID = en.PhotoExtractID
    JOIN xGroupAMIStoAMIS g1 ON g1.ObjID = n.ObjID
    JOIN xGroupAMIStoPrimary g2 ON g2.ObjID = f.ObjID
    LEFT OUTER JOIN xSdssDR6 gs
        ON gs.ObjID = n.ObjID AND gs.MatchMode = 1
WHERE gg.Distance BETWEEN 3 AND 5
    AND n.Band = 1 AND f.Band = 2 AND n.Mode = 1
    AND g1.N = 1 AND g2.N = 1 -- One-to-one GALEX match
\end{verbatim}
\end{minipage}
\end{center}
\caption{Revised band-merging in SQL; see text for details.}
\label{fig:sql-master}
\end{figure*}

In this analysis, we started with the official GALEX pipeline sources,
because it is the reference dataset that all studies rely on. Some of
these results have been already incorporated in the next GR4 release,
which should show improvements and will have to be studied further.
Our
immediate future work is to treat the GALEX FUV and NUV detections as
separate catalogs for the purpose of cross-identification with SDSS
and instead of performing a 2-way join between GALEX and SDSS, where
GALEX catalog is already the result of a previous 2-way match. This
new 3-way cross-matching of the SDSS and the two GALEX bands will be
done using the probabilistic formalism of
\citet{budavari_szalay}, however, for this next generation analysis
one will need a better understanding of the astrometry of the GALEX
sources, especially in the FUV, where the measurements are limited by
the small number of photons.

\acknowledgements

GALEX is a NASA Small Explorer.
We acknowledge NASA's support for construction, operation, and science
analysis for the GALEX mission, developed in cooperation with the
Centre National d'Etudes Spatiales of France and the Korean Ministry
of Science and Technology.
This research has made use of data obtained from and software provided
by the US National Virtual Observatory, which is sponsored by the
National Science Foundation.
T.B. gratefully acknowledges support from the
Gordon and Betty Moore Foundation via GBMF 554.


\end{document}